\documentclass[aps,prb,10pt,twocolumn,floatfix,superscriptaddress,showpacs,numerical,footinbib]{revtex4-1}

\usepackage{graphicx}
\usepackage{amsmath}
\usepackage{amssymb}
\usepackage{soul}

\usepackage[pdftex]{color}

\newcommand{\di}{\mathrm{d}}

\begin{document}

\title{Visualizing the chiral anomaly in Dirac and Weyl semimetals with photoemission spectroscopy}

\author{Jan Behrends}
\affiliation{\mbox{Max-Planck-Institut f\"ur Physik komplexer Systeme, N\"othnitzer Str.\ 38, 01187 Dresden, Germany}}

\author{Adolfo G.\ Grushin}
\affiliation{\mbox{Max-Planck-Institut f\"ur Physik komplexer Systeme, N\"othnitzer Str.\ 38, 01187 Dresden, Germany}}

\author{Teemu Ojanen}
\affiliation{\mbox{Low Temperature Laboratory, Department of Applied Physics, Aalto University, FI-00076 AALTO, Finland}}

\author{Jens H.\ Bardarson}
\affiliation{\mbox{Max-Planck-Institut f\"ur Physik komplexer Systeme, N\"othnitzer Str.\ 38, 01187 Dresden, Germany}}

\begin{abstract}
Quantum anomalies are the breaking of a classical symmetry by quantum fluctuations. 
They dictate how physical systems of diverse nature, ranging from fundamental particles to crystalline materials, respond topologically to external perturbations, insensitive to local details.
The anomaly paradigm was triggered by the discovery of the chiral anomaly that contributes to the decay of pions into photons and influences the motion of superfluid vortices in $^3$He-A.
In the solid state, it also fundamentally affects the properties of topological Weyl and Dirac semimetals, recently realized experimentally in TaAs, Na$_3$Bi, Cd$_3$As$_2$, and ZrTe$_{5}$.
In this work we propose that the most identifying consequence of the chiral anomaly, the charge density imbalance between fermions of different chirality induced by non-orthogonal electric and magnetic fields, can be directly observed in these materials with the existing technology of photoemission spectroscopy.
With angle resolution, the chiral anomaly is identified by a characteristic note-shaped pattern of the emission spectra, originating from the imbalanced occupation of the bulk states and a previously unreported momentum dependent energy shift of the surface state Fermi arcs. 
We further demonstrate that the chiral anomaly likewise leaves an imprint in angle averaged emission spectra, facilitating its experimental detection.
Thereby, our work provides essential theoretical input to foster the direct visualization of the chiral anomaly in condensed matter. 
\end{abstract}

\maketitle

%------------------
%-- Introduction --
%------------------

Quantum anomalies occur when a classically preserved symmetry is broken upon quantization~\cite{B96}.
The chiral anomaly, the first known example, refers to the non-conservation of a chiral current --- a current imbalance between two distinct species of chiral fermions.
Its discovery resolved the discrepancy between the measured and calculated decay rate of the neutral pion into two photons~\cite{B96}, and in condensed matter it was first identified as being responsible for a transverse force acting on vortices in superfluid $^3$He-A~\cite{V03,Volovik2013}.
In the solid state, Weyl and Dirac semimetals---TaAs was recently theoretically predicted~\cite{Weng2015,Huang2015} and experimentally identified using angle-resolved photoemission spectroscopy~\cite{SIN15,Lv2015} (ARPES) as being a Weyl semimetal, with Na$_{3}$Bi~\cite{Xu2013,Liu2014a,SJB15,Xu2015}, Cd$_3$As$_2$~\cite{Neupane2014,Borisenko2014,Yi2014,Liu2014b,Liang2015,He2014,Feng2014} and ZrTe$_{5}$~\cite{Li2014} likewise identified as Dirac semimetals---are described at low energies by chiral fermions that realize a chiral anomaly and host topological surface states with Fermi arcs~\cite{TV13,Hosur2013}.
In this work we theoretically explore the spectroscopic signatures of the chiral anomaly in these materials, with special emphasis on its effect on the Fermi arcs.  
We reveal an identifying note-shaped structure obtained in ARPES and a characteristic peaked structure of angular averaged photoemission spectra that allow for an experimental visualization of the chiral anomaly.

Weyl semimetals are a topological state of matter in which the conduction and valence bands touch and linearly disperse around pairs of Weyl nodes~\cite{TV13,Hosur2013}.
Each node has a definite left or right handed chirality providing a quantum number analogous to the valley degree of freedom in graphene~\cite{CastroNeto2009}.
Dirac semimetals can be thought of as two superimposed copies of Weyl semimetals with the degeneracy protected by a crystal symmetry from opening up a gap~\cite{Young2012,Wang2012,Wang2013,Weng2015,Huang2015}.
Similar to topological insulators and their metallic surface, Dirac and Weyl semimetals host protected surface states~\cite{Wan2011}
that unconventionally only exist for a restricted range of crystal momenta, thereby forming a Fermi arc connecting a pair of Weyl points with opposite chirality~\cite{Wan2011,Xu2011}.
Fermi arcs have been identified with ARPES in both TaAs~\cite{SIN15,Lv2015} and Na$_{3}$Bi~\cite{Xu2015}.

The chiral fermions describing the low energy degrees of freedom of Dirac and Weyl semimetals exhibit the chiral anomaly~\cite{Nielsen1983,Volovik:1999da,Aji:2012gs}:
while the sum of left and right handed fermions is necessarily conserved, their difference, the chiral density, does not have to be, even if classically it should.
In fact, non-orthogonal magnetic and electric fields pump left handed fermions into right handed, or vice versa\cite{Volovik:1999da,Aji:2012gs,ZyuzinBurkov2012,Grushin2012,Goswami:2013jp,Liu2013}.
Disorder induced internode scattering eventually counterbalances the pumping, leading to a non-equilibrium steady state with a nonzero chiral density.
The chiral anomaly is theoretically predicted to result in negative magnetoresistance~\cite{Abrikosov1998,Son2013} (experimentally observed in ZrTe$_{5}$~\cite{Li2014} and TaAs~\cite{Huang2015b,ZhangXu2015}), local~\cite{Landsteiner2014,Zhou2015} and non-local~\cite{Parameswaran2014} transport phenomena, chiral optical activity~\cite{Hosur2014,Ashby2014,Grushin2012} and rotation induced cooling~\cite{Basar2014}.
\begin{figure*}[tb]
 \includegraphics{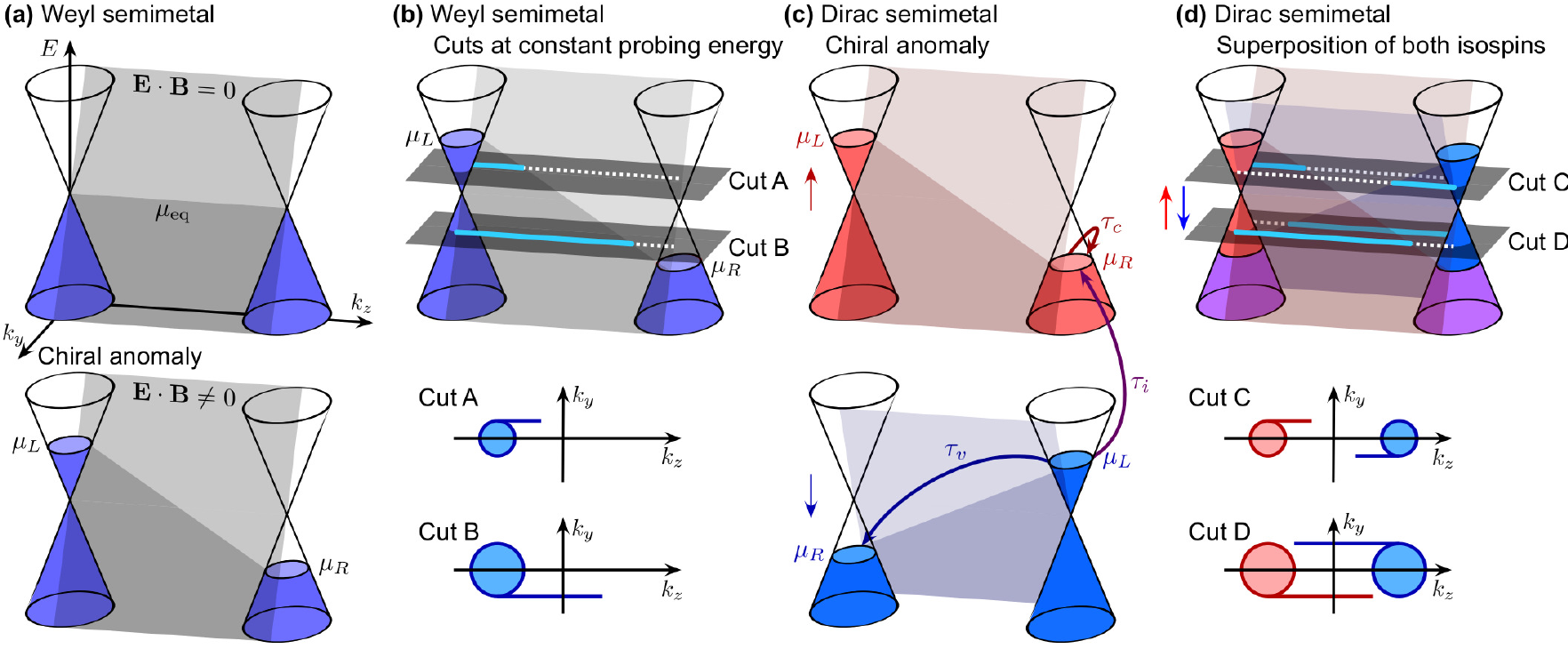}
 \caption{\textbf{Visualization of the chiral anomaly in Dirac and Weyl semimetals.}
  (a) Low-energy spectrum of a Weyl semimetal film with two bulk Weyl nodes of different chirality separated in momentum space. 
  The grey plane represents the surface state at the film's top surface, occupied up to the equilibrium chemical potential $\mu_\text{eq}$.
  Applying external magnetic and electric fields that satisfy $\mathbf{E} \cdot \mathbf{B} \neq 0$ results in a steady state with left and right cone chemical potentials $\mu_L \neq \mu_R$, linearly interpolated by a tilted Fermi arc.
  (b) Two constant energy cuts (A and B) through the band structure, with occupied and empty surface states depicted by solid light blue and white dashed lines respectively.
  The occupation at these cuts shows a characteristic blue note-shaped pattern depicted in the lower panel.
  (c) Dirac semimetals host pairs of Weyl cones, each pair with fixed isospin ($\uparrow$ or $\downarrow$) and both left and right chiralities, that respond to the chiral anomaly in the opposite way.
  Two edge states with opposite velocities (light red and light blue planes), appear at each boundary of the Dirac semimetal.
  Scattering processes within and between cones with scattering times $\tau_c$, $\tau_v$ and $\tau_i$ are depicted by arrows.
  (d) The two pairs of Weyl nodes in (c) together comprise a pair of Dirac nodes.
  At a fixed energy cuts (C and D) between $\mu_L$ and $\mu_R$, both bulk nodes are occupied while the surface states are only partially occupied. 
  The total occupation in these planes describes two facing note-shaped patterns, illustrated in the bottom panel.
  }
  \label{fig:chiral_anomaly}
\end{figure*} 

Photoemission spectroscopy (PES) has previously been overlooked as a diagnostic tool of the chiral anomaly despite its importance in observing topological semimetals.
A possible reason for this is that the finite magnetic field required to observe the chiral anomaly may complicate the disentangling of electron trajectories needed for angular resolution.
Setting aside this obstacle (we return to it at the end of this work where we argue that it can be overcome) we identify the main spectral signatures of the chiral anomaly and its observable effect on the Fermi arcs.
First, the bulk spectrum is determined by the differently occupied Weyl nodes distinguished by their chiralities;
second, the bulk chiral imbalance tilts the Fermi arcs, which then appear at fixed energy as finite segments stemming from the bulk Fermi surface.
Together, these two features form a distinguishing note-shaped photoemission pattern that we argue are within reach of current experimental state of the art.
We further calculate angle integrated PES, which does not suffer from the magnetic field complications of its ARPES relative, and show that distinct signatures of the chiral anomaly survive.
Overall, our results supply essential theoretical input that render photoemission spectroscopy a viable probe to visualize the chiral anomaly in Dirac and Weyl semimetals.
%

%-------------------------------------
% -- Properties of Weyl semimetals --
%-------------------------------------

To support our conclusions, we start by discussing the main features of the band structure of Weyl semimetals, shown in Fig.~\ref{fig:chiral_anomaly}a.
Their low-energy bulk spectrum consists of an even number of band touching points of left (L) and right (R) handed chirality that are separated in energy and momentum by breaking inversion or time-reversal symmetry.
Close to the Weyl points the energy dispersion is approximately linear and described by a  Weyl Hamiltonian
\begin{equation}
 \mathcal{H}_{L/R} = \pm \hbar v\,\boldsymbol\sigma \cdot \mathbf{k},
 \label{eq:weyl_equation}
\end{equation}
with $\mathbf{k}$ the momentum, $\boldsymbol\sigma$ the vector of Pauli matrices and $v$ the Fermi velocity.
Its eigenstates have a spin that points either radially away from or into the Weyl point, depending on its chirality. 
Each Weyl node therefore acts as a monopole of Berry flux in momentum space~\cite{Volovik2013,Vafek:2014hl}. 
In analogy with magnetic monopoles, a Dirac string necessarily emanates from the monopoles connecting a pair with opposite chirality~\cite{Dirac1931}.
Any two dimensional plane in momentum space, spanned say by the two momenta $k_x$ and $k_y$ at a fixed $k_z$, that crosses the Dirac string an odd number of times and does not contain a Weyl point defines a topologically nontrivial gapped band structure.
Therefore, if the system is made into a film that is finite in a real space direction conjugate to a momentum in that plane, for example either the $x$- or $y$-direction, a chiral surface state is obtained at the corresponding surfaces~\cite{Wan2011}.
Since $k_z$ remains a good quantum number we can repeat this arguments for different planes at different $k_z$. 
Only those crossing the Dirac string odd times have a surface state, which therefore only exists for certain values of the $k_z$ momenta.
The surface state dispersion is depicted as the grey shaded plane leaning on the two Weyl nodes in Fig.~\ref{fig:chiral_anomaly}a (the opposite surface provides an analogous surface plane with opposite velocity that is not shown).
The separator between occupied and unoccupied surface states is an arc --- the Fermi arc.

\begin{figure*}[tb]
 \includegraphics{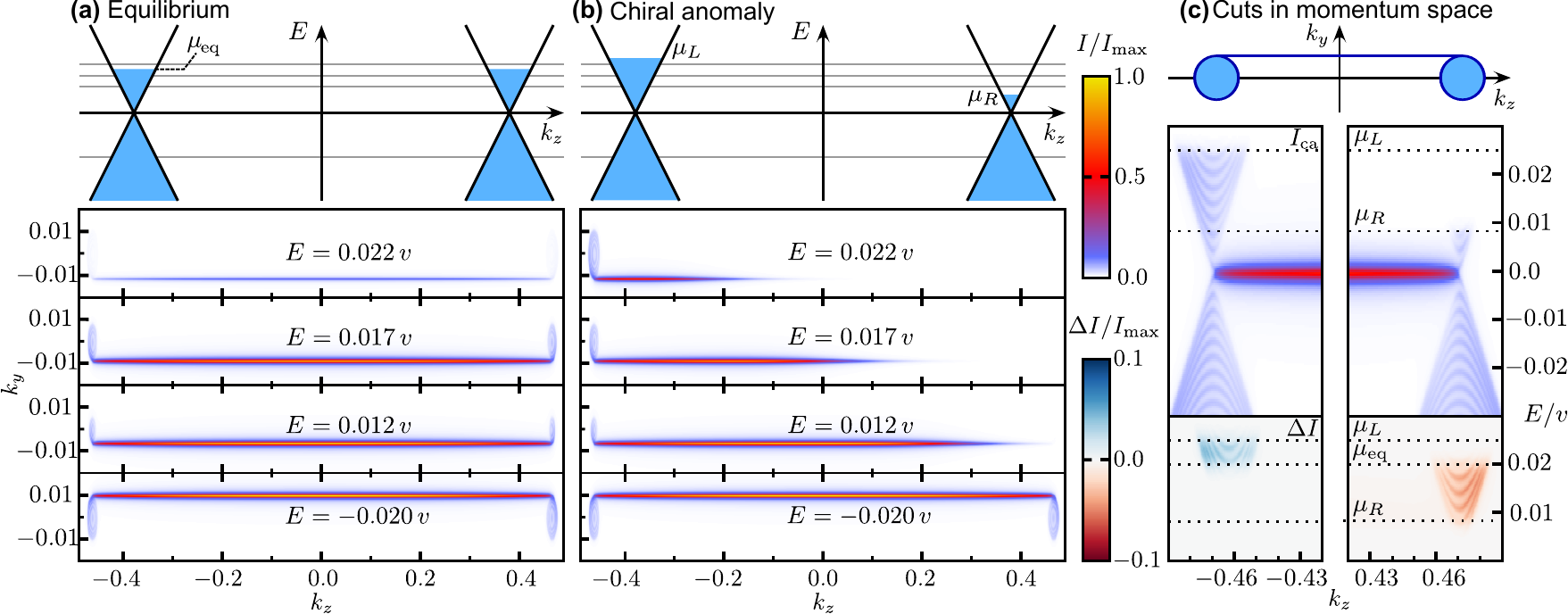}
 \caption{\textbf{ARPES signatures of the chiral anomaly in Weyl semimetals.} 
 Numerically computed ARPES spectra for a doped Weyl semimetal film with $L=2000$ layers in (a) equilibrium and (b) and (c) the chiral anomaly induced steady state.
 The parameters are such that the equilibrium chemical potential is $\mu_\mathrm{eq} = 0.02\,v$ while in non-equilibrium the left and right cones are filled up to 
 $\mu_L = 0.025\,v$ and $\mu_R = 0.008\,v$ respectively. 
 The lower panels in (a) and (b) show the momentum resolved spectra at a fixed energy, located as schematically shown with grey lines in the upper panels.
 In equilibrium (a), bulk cones and the surface state are observed, while the chiral anomaly (b) results in the disappearance of one cone and the emerging of the characteristic note-shaped pattern between $\mu_L$ and $\mu_R$. 
 The panel (c) shows the ARPES spectrum in non-equilibrium at a fixed $k_y = 0$, depicted in the top panel.
 The upper plot in the lower panel shows the total ARPES spectra, demonstrating the cone-like structure from the bulk and the flat surface state; the lower plot displays the intensity difference $\Delta I = I_\mathrm{ca} - I_\mathrm{eq}$ between the out-of-equilibrium and the equilibrium states.
 The dashed lines mark the chemical potentials $\mu_L$, $\mu_R$ and $\mu_\mathrm{eq}$. 
 These plots were obtained for $v = v_z$, $t=  0.5\,v$, $\epsilon = 6\,t$ and $b_z = 0.9\,v$ with the temperature set to $T=0.001v$, which corresponds to $T\approx 5$ K for a typical Fermi velocity of $v=0.45$ eV (see methods and supplementary material for details).
 }
 \label{fig:energy_cut_weyl}
\end{figure*}

%---------------------------------------
% - Chiral Anomaly in Weyl semimetals --
%---------------------------------------

In the presence of non-orthogonal external electric ($\mathbf{E}$) and magnetic ($\mathbf{B}$) fields, the chiral anomaly leads to a non-conservation of the left and right handed electron densities.
This is expressed by the two coupled continuity equations
\begin{equation}
\label{eq:cont}
  e\,\partial_t n_{L/R} + \nabla \cdot \mathbf{j}_{L/R} = \mp \frac{e^3}{4\,\pi^2\,\hbar^2} \mathbf{E} \cdot \mathbf{B}  \pm \frac{e}{2\,\tau_v}(n_R-n_L),
\end{equation}
where $n_{L/R}$ is the density of left and right handed fermions measured from the Weyl point and $\mathbf{j}_{L/R}$ their current density.
The first term on the right hand side is the anomaly contribution, which has a different sign for the two chiralities;
the second term represents the inter-valley scattering with rate $\tau_v^{-1}$.
At long times, a steady state with occupation difference between the two chiralities is obtained.
The continuity equations~\eqref{eq:cont} and particle conservation then define the left and right handed chemical potentials
\begin{equation}
 \mu_{L/R} = \left[\mu^3_{eq}\mp\frac{3}{2} \hbar\,v^3\,e^2\,\tau_v \,\mathbf{E} \cdot \mathbf{B}\right]^{1/3},
 \label{eq:mu_ca}
\end{equation}
where $\mu_\text{eq}$ is the equilibrium chemical potential and we have used that $n_{L/R} = \mu_{L/R}^3/(6\,\pi^2\,\hbar^3\,v^3)$ for three dimensional Weyl fermions.
In defining $\mu_{L/R}$ we assume that the equilibration\footnote{The definition of the relaxation times includes energy relaxation processes at small momenta, which we assume happen on a timescale shorter than any elastic scattering.} within a node with intra-valley relaxation time $\tau_c$ is much shorter than the inter-valley equilibration time $\tau_v$\cite{Parameswaran2014}.
This similarly suggests that relaxation along the surfaces is dominated by small momentum scattering such that the Fermi arc linearly interpolates in momentum space between $\mu_L$ and $\mu_R$, analogous to the voltage drop along an ohmic wire, leading to the steady-state occupation shown in the bottom of Fig.~\ref{fig:chiral_anomaly}a.
At a fixed energy between $\mu_R$ and $\mu_L$ this results in the characteristic note-shaped occupation schematically shown in Fig~\ref{fig:chiral_anomaly}b for the two constant energy planes denoted cut A and cut B. 

%-------------------------------------
% -- Properties of Dirac semimetals --
%-------------------------------------

Dirac semimetals can be understood as two copies of Weyl semimetals, with each Dirac node composed of two Weyl fermions of opposite chirality.
The two copies are distinguished by their different total angular momentum\cite{Wang2013}, which can be captured by an isospin quantum number that we denote with  $\uparrow$ and $\downarrow$. 
As long as the crystal symmetry is not broken, the isospin remains a good quantum number.
This has two important consequences:
one, the two chiral fermions comprising a Dirac node are decoupled and therefore a gap does not open;
two, the Weyl nodes still act as monopoles in momentum space and their Dirac string connects monopoles with opposite chirality but the \textit{same} isospin.
Dirac semimetals therefore have two Fermi arcs with opposite velocity on each surface.

\begin{figure*}[tb]
 \includegraphics{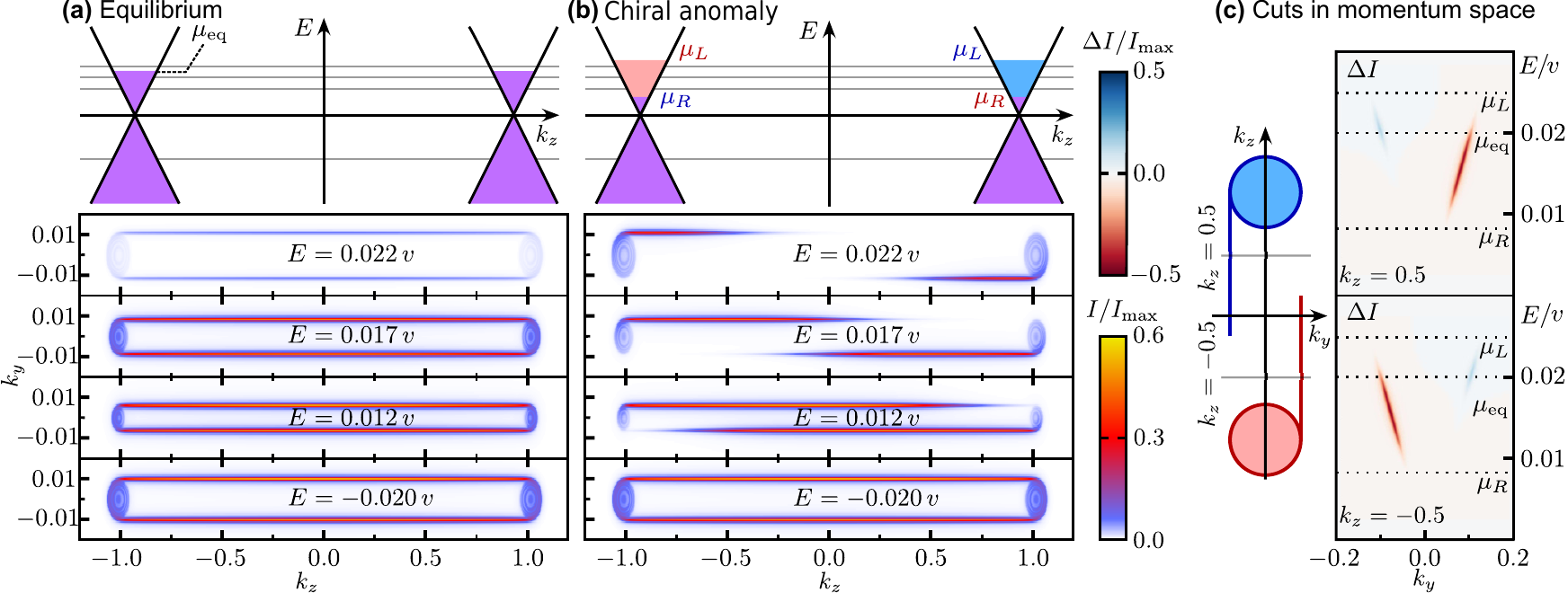}
 \caption{\textbf{ARPES signatures of the chiral anomaly in Dirac semimetals.} Numerically computed ARPES spectra for a doped Dirac semimetal film of $L=2000$ layers in (a) equilibrium and (b) and (c) the presence of $\mathbf{E}\cdot\mathbf{B}\neq0$.
 The spectra in (a) and (b) are the momentum resolved spectra at the fixed energies schematically shown at the top as gray lines. 
 The equilibrium chemical potential is set to $\mu_\mathrm{eq} = 0.020\,v$ while the left and right out of equilibrium chemical potentials are chosen to be $\mu_L = 0.025\,v$ and $\mu_R = 0.008\,v$.
 While there is a slight bulk Dirac cone intensity reduction from the equilibrium to the non-equilibrium situation, a stark qualitative difference is observed in the surface states that leads to the characteristic double note-shaped pattern.
 In (c) we plot the intensity difference $\Delta I = I_\mathrm{ca} - I_\mathrm{eq}$ for a fixed momentum $k_z = \pm 0.5$, which reveals the linear dispersion of the edge state.
 These plots were obtained for $M_0 = -0.2\,v$, $M_1 = -0.25\,v$, $M_2 = -0.75\,v$ and a temperature of $T = 0.001 \,v$.
 With $v=0.45\,\mathrm{eV}$ for Na$_3$Bi, the temperature corresponds to $T \approx 5\,\mathrm{K}$ and the induced chemical potential difference to $\delta \mu \approx 8\,\mathrm{meV}$.
 }
 \label{fig:energy_cut_dirac}
\end{figure*}
%

%----------------------------------------
% - Chiral Anomaly in Dirac semimetals --
%----------------------------------------

The two pairs of Weyl fermions in the Dirac semimetal are oppositely affected by the chiral anomaly.
External and non-orthogonal $\mathbf{E}$ and $\mathbf{B}$ fields shift the occupation in one Dirac cone to higher energies for one isospin and to lower for the other, and oppositely in the other Dirac cone, as schematically shown in Fig.~\ref{fig:chiral_anomaly}c.
For this steady state to be realized, intra-valley relaxation at fixed chirality must be larger than both the inter-valley relaxation, $\tau_c^{-1} \gg \tau_v^{-1}$, and the intra-valley relaxation between isospins, $\tau_c^{-1} \gg \tau_i^{-1}$ (the different relaxation processes are depicted with arrows in Fig.~\ref{fig:chiral_anomaly}c).
Both conditions are estimated to be satisfied in Dirac semimetals~\cite{Parameswaran2014}. 
Moreover, high mobility materials such as graphene satisfy the condition $\tau_c^{-1} \gg \tau_v^{-1}$, which allows for the experimental observation of a chemical potential imbalance~\cite{GSY14} and suggests that it is also met in Cd$_3$As$_2$.
The total occupation of a given Dirac cone is a superposition of both isospins, and therefore the bulk occupation in the chiral anomaly induced steady state is qualitatively the same as in its absence: two circular disks.
In contrast, the tilt of the Fermi arcs and the resulting partial occupation at a fixed energy, see Fig.~\ref{fig:chiral_anomaly}d, leads to a qualitatively new signature in the form of two facing note-shaped patterns (familiar from the Weyl semimetal case). 
This key property, which allows photoemission of Dirac semimetals to show evidence of the chiral anomaly, is a central result of our work.
%

%-----------------------------------------------------
%-- Revealing the Chiral Anomaly with ARPES -- Weyl --
%-----------------------------------------------------

We establish by a numerical computation that these note-shaped patterns are indeed directly manifested in photoemission spectra.
Our simulated data is obtained with exact diagonalization of tight binding models of Weyl and Dirac semimetals, as explained in the methods section.
In Fig.~\ref{fig:energy_cut_weyl}a we plot the momentum resolved ARPES spectra at various fixed energies for a doped Weyl semimetal in equilibrium.
Two bulk cones and one surface state at the probed surface are clearly seen; the surface state localized at the opposite surface is not visible due to the finite penetration depth of the incoming photon.
In the presence of external fields $\mathbf{E}\cdot\mathbf{B}\neq0$, the note-shaped pattern of the occupied states is seen in Fig.~\ref{fig:energy_cut_weyl}b for energies between $\mu_L$ and $\mu_R$.
An alternative way to illustrate the steady state occupation with $\mu_R \neq \mu_L$ is through a cut in momentum space at a fixed $k_y = 0$ as provided in Fig.~\ref{fig:energy_cut_weyl}c.
The pumping of charge between the different chiralities is clearly revealed in the intensity difference between the equilibrium $I_\mathrm{eq}$ and chiral anomaly induced steady state $I_\mathrm{ca}$, shown in the lower panel of Fig.~\ref{fig:energy_cut_weyl}c.
%

%-----------------------------------------------------
%-- Revealing the Chiral Anomaly with ARPES -- Dirac --
%-----------------------------------------------------
The numerically computed ARPES spectra for a Dirac semimetal, shown in Fig.~\ref{fig:energy_cut_dirac}, similarly reveal the essential features discussed in Fig.~\ref{fig:chiral_anomaly}.
In equilibrium two bulk cones and two counter-propagating edge states at the same surface are seen in Fig.~\ref{fig:energy_cut_dirac}a, similar to the experimental observations for Na$_3$Bi\cite{Xu2015}.
The non-equilibrium spectra in the presence of electric and magnetic fields are qualitatively different;
two copies of the note-shaped pattern clearly reveal the chiral anomaly in Dirac semimetals.
To further highlight this feature we plot, as for the Weyl semimetal, the intensity difference between the equilibrium and non-equilibrium in Fig.~\ref{fig:energy_cut_dirac}c for fixed $k_z = \pm 0.5$.
The qualitatively most notable feature is the partial occupation of the surface state that results in the stem of the note.
The bulk, though, is not entirely insensitive to the chiral anomaly:
first, states are occupied up to an energy $\max(\mu_L,\mu_R)$ that is higher than in the equilibrium situation;
second, only one isospin band is filled between $\mu_L$ and $\mu_R$ resulting in a decreased intensity.
From the chiral anomaly equation~\eqref{eq:mu_ca} we estimate the induced chiral chemical potential difference $\delta\mu=\mu_{L}-\mu_{R}$
to be within experimental state of the art.
For the doping levels of Na$_3$Bi, an electric field strength of $10^4\,\mathrm{V}\,\mathrm{m}^{-1}$ and a magnetic field of $1\,\mathrm{mT}$ gives $\delta \mu$ of the order of $10\;\textrm{meV}$, which is well within ARPES resolution.
Remarkably, magnetic fields as small as 6$\,\mu$T can still achieve observable $\delta \mu\approx 2.7\,$meV for Na$_3$Bi and can reach $\delta \mu\approx 5.4\,$meV in Cd$_{3}$As$_{2}$ 
(details of the experimental values used to compute these estimates for different materials as well as ARPES spectra specific to Na$_3$Bi with curved Fermi arcs, are given in the supplementary material).
The fact that experimental observation of Fermi arcs in equilibrium have already been reported also bodes well~\cite{SIN15,Lv2015,Xu2015}.
To achieve momentum resolution in ARPES it is necessary to correlate the angle at which an electron is detected to its initial momentum.
In free space, electrons move in straight lines making this task straightforward, and a constant electric field does not overly complicate it.
A magnetic field turns the electron trajectories into spirals and may, depending on the distance to the detector and size of the magnetic field, make momentum resolution difficult.
However, since the magnetic field strength needed for an observable effect is rather small, correcting induced deviations is plausible~\cite{E10}.
Alternatively, a larger uniform magnetic field finite within the sample could in principle be engineered with a ferromagnetic material in the picture frame geometry: a closed magnetic circuit that minimizes stray fields outside the sample that has been used to experimentally study ferromagnetic metals via ARPES~\cite{RHC83} (see supplementary material for details).
Similarly, if either field can be turned off fast enough (faster than the inter-valley relaxation) a pump-probe setup could observe the non-equilibrium steady state and its equilibration, which would allow for a direct measurement of the inter-valley relaxation time in addition to visualizing the chiral anomaly.
While none of these are simple tasks, we believe that the rewards are significant enough that the experimental challenge will be met.

\begin{figure}
 \includegraphics{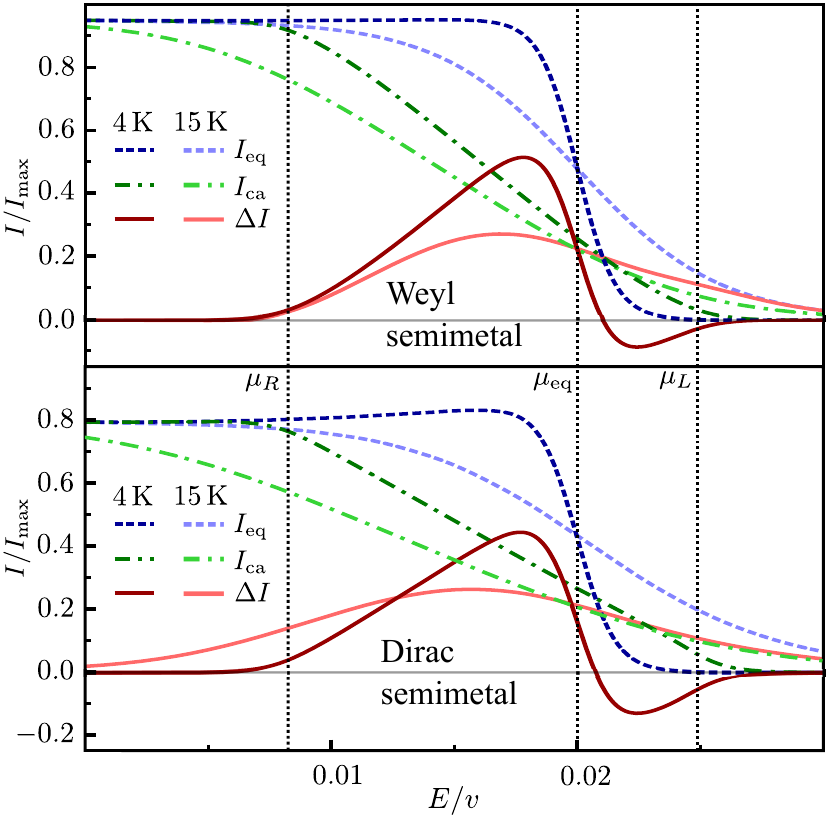}
 \caption{\textbf{Chiral anomaly in PES spectra for Dirac and Weyl semimetals.} Numerically computed PES intensities for Weyl and Dirac semimetals for two different temperatures.
 In equilibrium the intensity $I_{\mathrm{eq}}$ presents a single step around $\mu_\mathrm{eq}$, that is smeared out by increasing temperature and resembles 
 the occupation of the cones.
 In non-equilibrium the intensity $I_{\mathrm{ca}}$ shows a double-step-profile with steps at $\mu_R$ and $\mu_L$.
 Both $\mu_{R/L}$ include the temperature dependence of the chiral anomaly \cite{V03,Li2014}.
 The difference $\Delta I =I_{\mathrm{eq}}- I_{\mathrm{ca}}$ highlights the effect of the chiral anomaly by a characteristic peak-dip structure for low temperatures, resulting from the shift in occupation of the two cones, that evolves into a single peak as temperature is increased.
 The parameters used to obtain these plots are $\mu_\mathrm{eq} = 0.020\,v$, $\mu_L = 0.025\,v$, $\mu_R = 0.008\,v$ at $4\,$K and $\mu_L = 0.024\,v$, $\mu_L = 0.005\,v$ at $15\,$K.
 The temperatures are chosen to be $T=8\cdot 10^{-4}\,v$ and $T=2.9\cdot10^{-3}\,v$, corresponding to $4\,$K and $15\,$K for a typical value of $v=0.45\,$eV.
The rest of parameters are chosen as in Fig.~\ref{fig:energy_cut_weyl} and~\ref{fig:energy_cut_dirac} for the upper and lower panel respectively.
 }
 \label{fig:pes}
\end{figure}

In the absence of momentum resolution the angular averaged but energy resolved photoemission spectroscopy likewise contains direct signatures of the chiral anomaly.
In Fig.~\ref{fig:pes} we plot the numerically computed PES spectra as a function of energy for both Weyl and Dirac semimetals.
In equilibrium it has a single step that is smeared by temperature.
In the non-equilibrium steady state occupation is shifted from lower energies to higher, such that the bulk spectra would have a double step profile, with one step at $\mu_L$ and the other at $\mu_R$.
In the total spectra the two steps are hard to see since the surface states contribute significantly to smoothen the profile.
The intensity difference $\Delta I = I_\text{eq}-I_\text{ca}$ instead shows a characteristic peak-dip structure for low temperatures, which reflects the chiral anomaly pumping of fermions of one chirality into the other, evolving into a single peak as temperature is increased.

% ----------------------------
% -- Summary and Conclusion --
% ----------------------------

In conclusion, we have promoted photoemission as an ideal tool to experimentally visualize the chiral anomaly in the solid state.
The main effect that allows this is the tilt of the Fermi arc in the presence of non-orthogonal electric and magnetic fields that, through the chiral anomaly, pump electrons from one chirality to the other.
As a consequence, the surface state occupation at a fixed energy no longer connects bulk nodes but rather terminates in between, resulting in a qualitatively distinct note-shaped photoemission pattern identifying the chiral anomaly.
We argued, by estimating from experimentally available parameters the relevant chiral chemical potential difference to be about $10\,\text{meV}$, that a direct visualization of the chiral anomaly is within the state of the art.
Were it to be realized experimentally in the way proposed here, a revealing light would be shed on the relation between two fundamental concepts: quantum anomalies and topological states of matter. 
\section*{Methods}
All presented ARPES spectra are calculated via exact diagonalization of tight binding Hamiltonians describing Weyl and Dirac semimetal films. 
For a Weyl semimetal we take
\begin{subequations}
\begin{align}
	& H_\text{WSM} = H_0 + H_1, \\
	& H_0 = 2\,v ( \sin k_y\,\sigma_x - \sin k_x\,\sigma_y)\,\tau_x + 2\,v_z \sin k_z\,\tau_y + M_\mathbf{k} \tau_x, \\
	& H_1 = b_0\,\sigma_z\,\tau_y + \mathbf{b} \cdot \left(-\sigma_x\,\tau_x, \sigma_y\,\tau_x,\sigma_z \right).
\end{align}
\end{subequations}
The first term models a three-dimensional topological insulator~\cite{Qi2011}.
The Pauli matrices $\boldsymbol\sigma$ and $\boldsymbol\tau$ respectively act in spin and particle-hole space.
The velocity $v$ in the $x$- and $y$-directions differs from that in the $z$-direction ($v_z$) consistent with experimentally relevant materials.
The mass $M_\mathbf{k} = \epsilon - 2\,t\,\sum_i \cos k_i$.
At $\epsilon = 6\,t$ a degenerate three-dimensional Dirac cone is obtained at $\mathbf{k} = 0$.
The second term\cite{Vazifeh2013} breaks inversion with a nonzero $b_0$ and time-reversal symmetry with a finite $\mathbf{b}$. 
In our simulation we take $b_{0}=0$ and fix the direction of $\mathbf{b}=(0,0,b_{z})$ such that the Dirac cone is split into two Weyl cones at $k_z \approx \pm \frac{b_z}{2\,v_z}$.
The low-energy spectrum of both Dirac semimetals Na$_3$Bi\cite{Wang2012} and Cd$_3$As$_2$\cite{Wang2013} is modeled by the Hamiltonian
\begin{equation}
 H_\text{DSM} = M_\mathbf{k} \,\sigma_z + 2\,v\,(\sin k_x \,\sigma_x\,\tau_z - \sin k_y\,\sigma_y ),
 \label{eq:dirac_hamiltonian}
\end{equation}
which can be understood as two copies of two-band Weyl semimetal Hamiltonians.
The mass term $M_\mathbf{k} = M_0 + 2\,M_1 (\cos k_z -1) + 2\,M_2 (\cos k_x + \cos k_y -2)$ sets the velocity in $k_z$ direction to $v_z = 2\,\sqrt{M_0\,M_1} $ and the Dirac cones to be at $k_z^c = \pm \sqrt{M_0/M_1}$. 
In this simple model the surface states are dispersionless in the $k_z$-direction.

To model a film we take the system finite in the direction orthogonal to the separation of the cones in momentum space, which is the surface where the edge states are maximally visible. 
The ARPES spectrum is given by an integral over the local density of states
\begin{equation}	
 I (\mathbf{k}_\parallel,\omega ) = \int \di x  \sum_n |\psi_{n,\mathbf{k}_\parallel} (x)|^2\,\delta ( \epsilon_{n,\mathbf{k}_\parallel} - \omega)  w(x)  f (\omega - \mu_{\mathbf{k}_\parallel} ),
\end{equation}
where $x$ is the coordinate in the finite direction of the film, $w(x) = \exp(-x/\ell)$ is a weight function modelling the incoming light's intensity decay with depth into the sample with decay length $\ell$, and $f$ is the Fermi-Dirac distribution.
The wavefunctions $\psi_{n,\mathbf{k}_\parallel}$ are the eigenfunctions of the Hamiltonians and $\epsilon_{n,\mathbf{k}_\parallel}$ the corresponding eigenvalues, which depend on the momentum $\mathbf{k}_\parallel$  parallel to the surface.
To model the chiral anomaly induced steady state, the chemical potential $\mu_{\mathbf{k}_\parallel}$ is taken to be $\mathbf{k}_\parallel$- and isospin-dependent. 
For the surface states, $\mu_{\mathbf{k}_\parallel}$ depends linearly on $k_z$.\\
\section*{Acknowledgements}
We thank Sergey Borisenko, Bernd B\"{u}chner and Ilya Belopolski for useful discussions about Dirac semimetals and possible extensions of current ARPES setups,
and S. Bera for helpful insights on numerical methods.

%\bibliography{database.bib}
%\bibliographystyle{naturemag}

%

\appendix

\section{\label{sec:aspects}Aspects of experimental implementation}

\subsection{\label{subsec:relax}Relaxation rates}
The experimental feasibility of detecting the chiral anomaly with photoemission spectroscopy relies firstly on the correct hierarchy of the different relaxation rates involved.
As discussed in the main text, for Weyl semimetals the intra-valley relaxation rate at fixed chirality must be faster than the inter-valley relaxation rate, $\tau_c^{-1} \gg \tau_{v}^{-1}$.
For Dirac semimetals it is also required that the relaxation rate between the two isospins forming each Dirac node must be shorter than the intra-valley relaxation, $\tau_i \gg \tau_c$.
The intra-valley relaxation relaxation rate $\tau_{c}$ can be deduced from the experimental values for the carrier mobilities $\mu_e$ given in the Table~\ref{tab:experiment} using that 
\begin{equation}\label{eq:mob}
\mu_e=\dfrac{\sigma}{e n} = \dfrac{e}{\hbar}\dfrac{v_{F}}{k_{F}}\tau_{c},
\tag{S\arabic{equation}}
\stepcounter{equation}
\end{equation}
where $\sigma$ denotes the DC conductivity, $e$ is the electric charge, and $n$ is the carrier density.  
A theoretical estimate of the ratio $\tau_v/\tau_c$ can be calculated in the first Born approximation following Ref.~\onlinecite{Parameswaran2014},
rendering the values included in the Table~\ref{tab:experiment} that justify the assumptions used in the main text.

Experimentally, $\tau_v$ can be determined via non-local transport measurements~\cite{Parameswaran2014}.
The corresponding inter-valley scattering length $\ell_v$ was obtained experimentally in Ref.~\onlinecite{Zhang2015b} for Cd$_3$As$_2$.
This length is connected to the scattering time via $\ell_v= \sqrt{ D\,\tau_v}$, where $D = \mu_e\,k_B\,T/e$ is the charge diffusion coefficient at temperature $T$.
Together with mobility measurements, the inter-valley scattering time $\tau_v$ at $T = 4\,\mathrm{K}$ can be determined to be $\tau_v \sim 10^{-9}\,\mathrm{s}$, similar to the theoretical estimate for Na$_3$Bi.

\begin{table*}
 \begin{tabular}{c|lc|lc|lc}
 \toprule
 & \multicolumn{4}{c|}{\textbf{Dirac Semimetals}} & \multicolumn{2}{c}{\textbf{Weyl Semimetal}} \\
 \cline{2-7}
  ~ & Cd$_3$As$_2$ &Ref. &  Na$_3$Bi & Ref. & TaAs & Ref.\\
  \toprule
  $\mu_e~(\mathrm{cm}^2\,\mathrm{V}^{-1}\,\mathrm{s}^{-1})$
	& $10^4$ -- $10^5$	& \onlinecite{JayGerin1977,Neupane2014,He2014,Feng2014}	& $\sim 10^3$--$10^4$	& \onlinecite{Xiong2015}\footnote{ Ilya Belopolski, private communication.}	& $5\,\cdot 10^5$	& \onlinecite{Zhang2015} \\
  \colrule
  $\tau_c~(\mathrm{s})$~[Eq. \eqref{eq:mob}]	&  $\sim10^{-14}$--$10^{-13}$ & &  $\sim10^{-13}$ & &  $\sim10^{-12}$ & \\
  \colrule
  $\tau_v / \tau_c$ \hspace{1mm} [theoretical]
	& -			& 				& $10^4$		& \onlinecite{Parameswaran2014}	& $500$		& \onlinecite{Parameswaran2014} \\
  \colrule
  $\tau_i / \tau_c$ \hspace{1mm} [theoretical]
	& -			& 				& $10^3$		& \onlinecite{Parameswaran2014}	& \multicolumn{2}{c}{$\tau_{i}$ not defined} \\
  \colrule
  $\tau_v~(\mathrm{s})$~[experiment]
	& $10^{-9}$		& \onlinecite{Zhang2015b}\footnote{Measurement of $\ell_v$, the inter-valley relaxation time was obtained as described in the text.}	& -			& ~				& -		& ~ \\
  \colrule
  $v~(\frac{\mathrm{m}}{\mathrm{s}})$
	& $7.6 \cdot 10^5$	& \onlinecite{Borisenko2014}	& $3.74 \cdot 10^5$	& \onlinecite{Liu2014a} & $3.1$ -- $3.6 \cdot 10^5$ (W1) &  \onlinecite{Huang2015b}\\
	& $9.3 \cdot 10^5$ 	& \onlinecite{Liang2015}	& & &  $2.6$ -- $4.3 \cdot 10^5$ (W2) &  \onlinecite{Huang2015b} \\
	& $1.1 \cdot 10^6$	& \onlinecite{He2014}		& & & & \\
	& $1.3 \cdot 10^6$	& \onlinecite{Liu2014b}		& & & & \\
	& $1.5 \cdot 10^6$	& \onlinecite{Neupane2014}	& & & & \\
  \colrule
  $v_z~(\frac{\mathrm{m}}{\mathrm{s}})$
	& $\sim 10^5$		& \onlinecite{Neupane2014}	& $2.89 \cdot 10^5$	& \onlinecite{Liu2014a}	& $3.4\cdot10^5$ (W1) & \onlinecite{Huang2015b} \\
	& $3.3 \cdot 10^5$	& \onlinecite{Liu2014b}		& & & $4.1\cdot10^4$ (W2) & \onlinecite{Huang2015b} \\
 \botrule
 \end{tabular}
\caption{Summary of experimentally measured Fermi velocities $v$ and $v_z$, mobilities $\mu_e$, and theoretical estimates for the relevant scattering times. For TaAs
W1 (W2) nomenclature classifies the eight (sixteen) Weyl points falling on (away) the $k_{z}=2\pi/c$ plane. The scattering rate $\tau_{c}$ is estimated for an isotropic Weyl or Dirac cone with $E_{\mathrm{F}}=\hbar v k_{F}\sim10\,$meV.}
\label{tab:experiment}
\end{table*}
\subsection{ARPES in finite magnetic fields}
ARPES experiments in magnetic fields are challenging. 
External magnetic fields affect electron trajectories (especially those with low energies) and compromise angle resolution.
Typical ARPES equipment is protected from external magnetic fields using
$\mu$-metal shields made from a metallic alloy with high magnetic permeability $\mu$. 
This material offers the magnetic field lines a path with a low magnetic resistance (or reluctance) that is inversely proportional to $\mu$, preventing them going to the energetically costly exterior. 
In typical ARPES experiments, for instance the set-up of Ref.~\onlinecite{E10}, high resolution requires the field inside 
the ARPES lens to be of the order of $\mu$T or less.
In practice up to $\sim6\,\mu$T can be handled~\cite{E10}. 
These small magnetic fields typically induce a rigid shift of the electron trajectories that can in principle be corrected~\cite{E10}.
As we discuss in the next section, already these small fields can result in observable values of the chiral potential difference $\delta\mu\sim 4\,$meV.

\begin{figure}
\begin{center}
\includegraphics[scale=0.25]{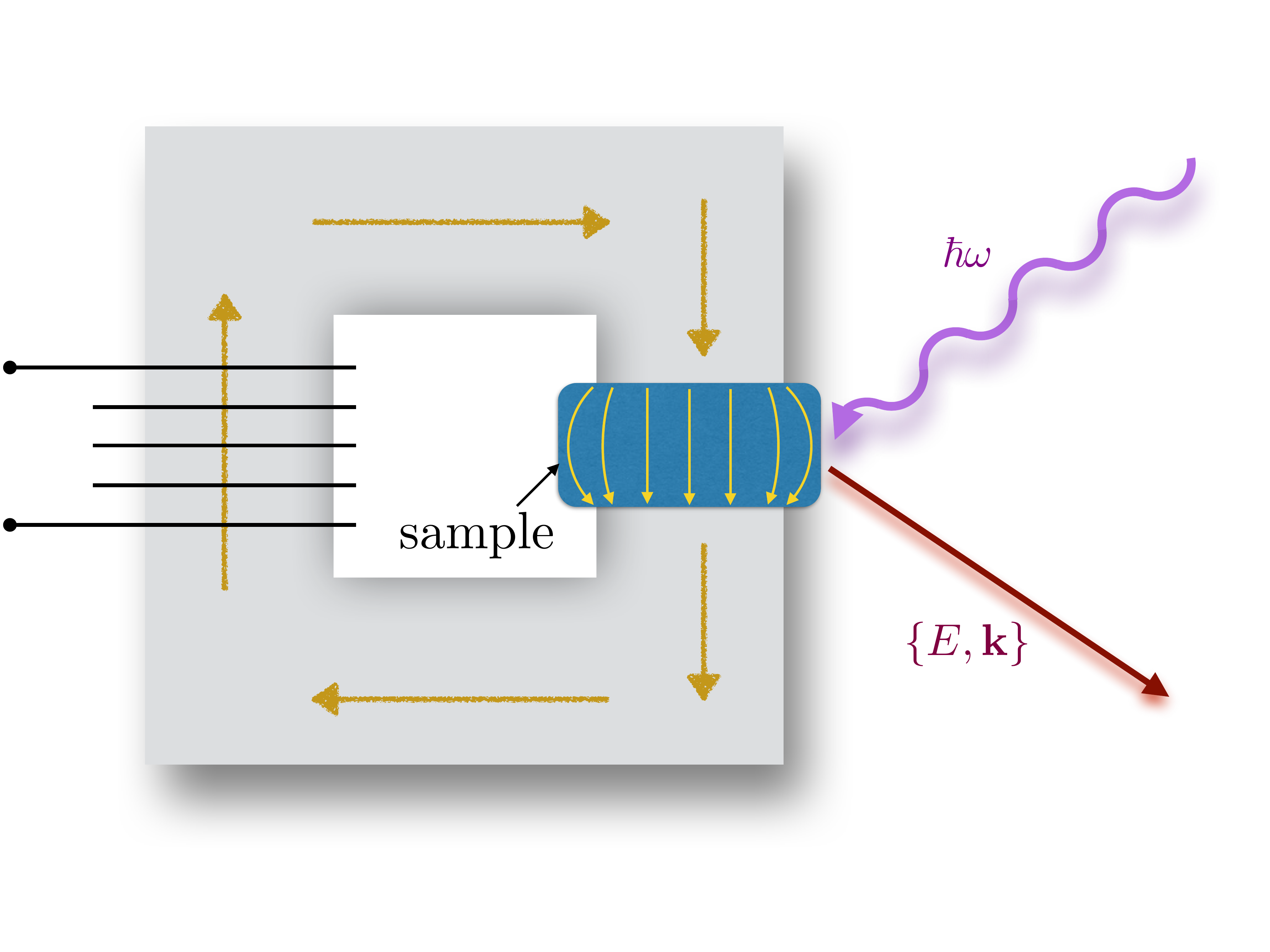}
\caption{\label{Fig:magcirc} Proposed magnetic circuit for studying the chiral anomaly and ARPES in a magnetic field, based on ARPES experiments on ferromagnetic materials~\cite{RHC83}. 
The sample is included in a magnetic circuit to minimize stray fields effects on photo-electron trajectories. 
The yellow arrows show the confined magnetization in the ferromagnetic material (e.g. nickel or iron) shaped into the picture-frame geometry.
The Weyl or Dirac semimetal sample, a photon of energy $\hbar\,\omega$ and a photo-electron with energy and momentum $\left\lbrace E, \mathbf{k} \right\rbrace$ are depicted schematically 
as a blue rectangle, a purple curved arrow and a red straight arrow respectively.}
\end{center}
\end{figure}

To allow for measurements in larger magnetic fields, without altering electron trajectories, it would be advantageous to go one step beyond existing set-ups.
We suggest that a possible experimental design can be based on early ideas used to perform ARPES on ferromagnetic materials~\cite{RHC83,D94}.
Stray fields around ferromagnetic materials, like nickel, can severely alter photo-electron trajectories and jeopardize ARPES measurement accuracy.
An elegant experimental solution studied these materials in the so-called picture frame geometry or remnant state~\cite{RHC83,D94} (see Fig.~\ref{Fig:magcirc}).
In this geometry, the magnetic field lines are confined within the material in a magnetic loop, minimizing their effect on electron trajectories that can spoil ARPES measurements. 
To reach a uniform field within a Weyl or Dirac semimetal sample, while minimizing the external stray fields, one possible solution is to open up a small gap in the magnetic circuit as shown in Fig.~\ref{Fig:magcirc}. 
In this set-up, a very uniform field is obtained within the gap, with possible additional fringing fields depicted as curved field lines.
The effect of the fringing field is to increase the effective area of the gap,
thus reducing its reluctance and increasing the effective magnetic field felt by the sample~\cite{T12}. 
The magnitude and spatial extent of the fringing field can therefore be optimized by increasing the gap's length or cross section.
In addition, if the sample cross section is designed to be slightly larger than the cross section of the gap, the fringing fields could
enter the sample as well, not affecting photo-electron trajectories as shown in Fig.~\ref{Fig:magcirc}.
Finally we note that external current loops were commonly used in experiments to orient the magnetization~\cite{RHC83,D94}.
Therefore it is plausible that these set-ups can admit modifications to incorporate electric fields as well.
\subsection{Estimates of the chiral chemical potential difference}

To observe the note-shaped pattern that signals the chiral anomaly it is essential that the induced chemical potential difference $\delta\mu$ is larger than the energy resolution of the experiments, which is of the order of $\sim\,\mathrm{meV}$~\cite{Liu2008} if magnetic fields are kept of the order of $\sim\mu$T.
To estimate the difference $\delta \mu=\mu_{L}-\mu_{R}$, we use the definition of the chiral chemical potentials given by equation (3) of the main text.
For the particular case where $\mu_\mathrm{eq} = 0$, this difference is given by
\begin{equation}
\label{seq:chiraldiff}
\delta \mu = 2\left( \dfrac{3}{2}\, \hbar\,v^2\,v_z\,e^2\,\tau_v \,\mathbf{E} \cdot \mathbf{B}\right)^{1/3},
\tag{S\arabic{equation}}
\stepcounter{equation}
\end{equation}
where $v$ and $v_z$ are the anisotropic cone Fermi velocities defined in the Methods section of the main text and reproduced in the Table~\ref{tab:experiment} as measured by ARPES for Na$_3$Bi and Cd$_{3}$As$_2$.
From Eq.~\eqref{seq:chiraldiff} we observe that it is the magnitude of the Fermi velocities, rather than $\tau_{v}$ or $\mathbf{B}$, that can most effectively enhance the chiral anomaly.
\\
To calculate $\delta\mu$ we choose two different magnetic field values $|\mathbf{B}|=6\,\mu$T and  $|\mathbf{B}|=1\,\mathrm{mT}$. 
The first corresponds to a reasonable magnetic environment that can be achieved in ARPES as discussed in the previous section, while the second
is a very conservative estimate of the magnetic fields that can be achieved with the frame geometry also introduced above. 
All our following estimates are calculated for $|\mathbf{E}| = 1\cdot10^4\,\mathrm{V}\,\mathrm{m}^{-1}$.
For the relaxation rates, we choose the experimentally obtained time $\tau_{v}=10^{-9}\,\mathrm{s}$ for Cd$_3$As$_2$ and the theoretical estimate $\tau_{v}/\tau_{c}=10^{4}$ for Na$_3$Bi, both justified in Sec.~\ref{subsec:relax}.
These are also conservative estimates since $\tau_{v}$ is expected to be even larger for small chemical potentials~\cite{Parameswaran2014}.
Given these values, we find for Cd$_3$As$_2$ and field strengths of $|\mathbf{B}|=1\,\mathrm{mT}$~($6\,\mu$T) a chiral chemical potential difference of $\delta \mu \sim 30 \,\mathrm{meV}$~($5.4 \,\mathrm{meV}$).
For Na$_3$Bi, we find that $\delta \mu \sim 15 \,\mathrm{meV}$~($2.7 \,\mathrm{meV}$) for $|\mathbf{B}|=1\,\mathrm{mT}$~($6\,\mu \mathrm{T}$).

\vspace*{0.25cm}
\section{Case example: $\mathrm{Na}_3\mathrm{Bi}$}
\begin{figure*}[tb]
 \includegraphics{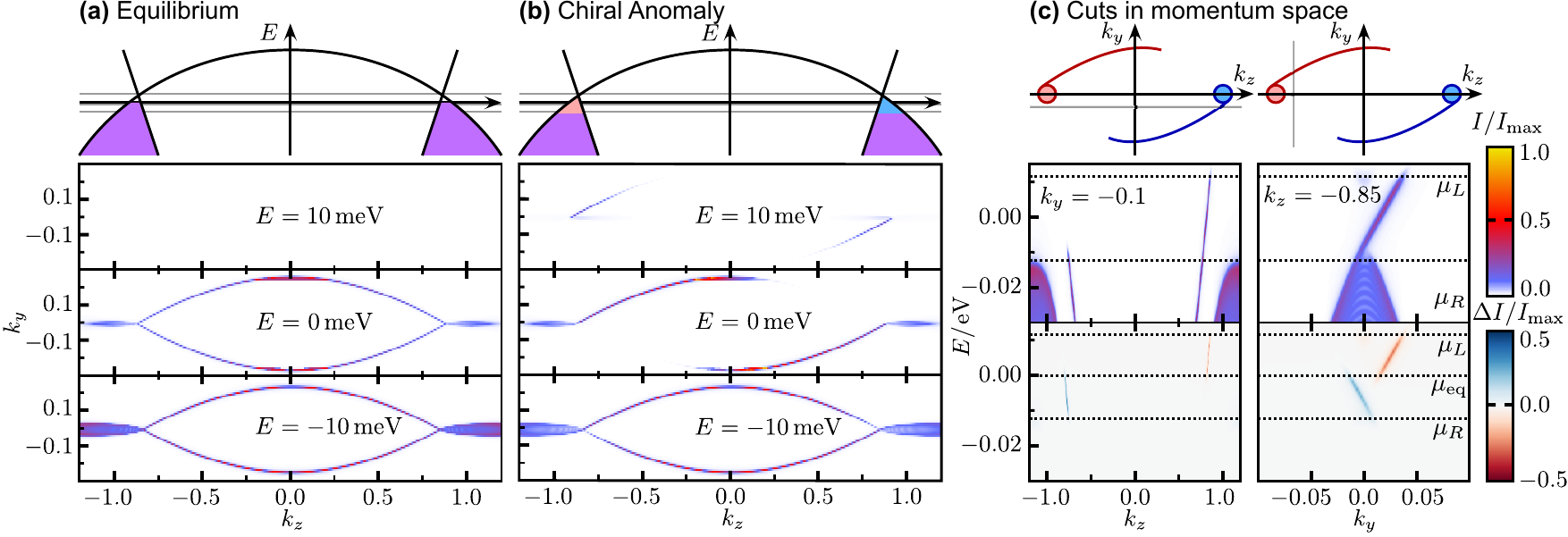}
 \caption{\textbf{Visualization of the chiral anomaly in Na$_3$Bi}
 (a) Numerically computed equilibrium ARPES spectra for a doped Dirac semimetal film Na$_3$Bi with $\mu_{\mathrm{eq}}= 0$.
 The pair of Fermi arcs are well within experimental energy and momentum resolution, confirmed by the recent experiment Ref.~\onlinecite{Xu2015}.
 (b) Chiral anomaly induced ARPES spectra for the same material and an estimate of the chiral chemical potential difference $\delta\mu \approx 20\,\mathrm{meV}$ derived in section \ref{sec:aspects}.
 The Fermi arcs show evidence of partial occupation within experimental resolution. 
 The upper panels in (a) and (b) show a schematic representation of the bulk band structure with two Dirac nodes connected at higher energies. 
 (c) The lower panel shows two cuts through momentum space at $k_y = -0.1$ and $k_z = -0.85$ in units of the lattice constants and
  represented schematically by the horizontal and vertical light grey lines in the upper panel.
 The parameters of the low energy model used to obtain these figures are extracted from first principles in Ref.~\onlinecite{Wang2012} and take the values 
 $v = 0.45\,\mathrm{eV}$, $M_0 =-0.087\,\mathrm{eV}$, $M_1 = -0.11\,\mathrm{eV}$, $M_2 = -0.35\,\mathrm{eV}$ and $C_0 = -0.064\,\mathrm{eV}$, $C_1 =  0.094\,\mathrm{eV}$, $C_2 =  -0.28\,\mathrm{eV}$.
  All calculations were performed at a temperature of $T = 1\,\mathrm{meV} = 11.6\,\mathrm{K}$ and for a film thickness of $L=1000$ layers.
  }
  \label{fig:suppchiral_anomaly}
\end{figure*}

The main features that can lead to a direct visualization of the chiral anomaly in Weyl and Dirac semimetals via photoemission spectroscopy 
are presented in the main text.
In order to make definite contact with current experimental state of the art,
we discuss the features of the chiral anomaly for $\mathrm{Na}_3\mathrm{Bi}$.
To this extent, we combine the experimental constants in Table~\ref{tab:experiment} with parameters obtained from \emph{ab initio} calculations for this material 
that define the realistic low energy model described in Ref.~\onlinecite{Wang2012}.
The equilibrium ARPES spectra computed for such model, as detailed in the main text, is shown in Fig.~\ref{fig:suppchiral_anomaly}(a).
The figure shows a clear pair of curved Fermi arcs that connect two Dirac nodes.
Distinguishing clearly these pair of arcs is well within experimental resolution, as confirmed by recent experiments [cf. Fig.~2A of Ref.~\onlinecite{Xu2015}].
Upon applying external fields that satisfy $\mathbf{E}\cdot\mathbf{B}\neq0$, the chiral anomaly induces a chiral chemical
potential difference of the order $\delta\mu\sim 10\,\mathrm{meV}$, as estimated in the previous section.
In Fig.~\ref{fig:suppchiral_anomaly}(b), we show the chiral anomaly induced ARPES spectrum for Na$_3$Bi corresponding to such
value of $\delta\mu$.
The Fermi arcs are partially occupied and thus end before reaching the bulk Fermi surface, the latter not visible within this resolution.
The observation of such features would directly visualize the chiral anomaly and confirm its effect on the surface electronic spectrum 
predicted in the main text.

\end{document}